# A Threshold for Laser-Driven Linear Particle Acceleration in Unbounded Vacuum

Liang Jie Wong[1, a)] and Franz X. Kärtner[1,2]

We hypothesize that a charged particle in unbounded vacuum can be substantially accelerated by a force linear in the electric field of a propagating electromagnetic wave only if the accelerating field is capable of bringing the particle to a relativistic energy in its initial rest frame during the interaction. We consequently derive a general formula for the acceleration threshold of such schemes and support our conclusion with the results of numerical simulations over a broad range of parameters for different kinds of pulsed laser beams.

Energetic particle beams are crucial to the progress of fields across the spectrum of science and technology, from cancer treatment [1] to particle physics [2] to inertial confinement fusion [3], nanolithography [4] and radioactive waste management [5]. High-intensity laser systems, made possible by chirped-pulsed amplification [6], can provide accelerating gradients that surpass those of conventional accelerators by as much as six orders of magnitude [7], paving the way to an era of table-top particle accelerators and table-top x-ray laser systems. Although plasma-based acceleration schemes [8] have had much experimental success, the possibility of accelerating particles in vacuum [9, 10] remains of great interest since the absence of plasma would preclude problems associated with the inherent instability of laser-plasma interactions.

Among the many vacuum-based acceleration schemes proposed is the acceleration of particles (primarily) by a force linear in the electric field of the laser [11-19]. This scheme may be realized with an electromagnetic wave that vanishes completely on the beam axis except for its longitudinal electric field component. As a result, an on-axis charged particle experiences only a force along the axis. Particles that are slightly off-axis will experience some ponderomotive acceleration due to non-zero transverse field components, but the longitudinal linear force should dominate.

In this Letter, we obtain a formula for the threshold power of net linear acceleration (a.k.a. direct acceleration) in unbounded vacuum. We hypothesize that a charged particle (regardless of initial energy) in unbounded vacuum can be substantially accelerated by a force linear in the electric field of a propagating electromagnetic wave only if the accelerating field is capable of bringing the particle to a relativistic energy in its initial rest frame during the interaction. By "substantial acceleration" we mean the ratio of final to initial particle energy $\gamma_f / \gamma_0 \gg 1$. Based on our hypothesis, we derive a formula for the threshold power and compare the formula with the results of exact numerical simulations over a broad range of parameters for different kinds of pulsed laser beams. The accuracy with which the formula matches our numerical simulations lends credence to our hypothesis and sheds light on the physical mechanism that enables net linear acceleration in unbounded vacuum: namely, that the ability of the accelerating field to bring the particle to a relativistic energy in its initial rest frame is critical to substantial net linear acceleration.

We will always assume that the pulse starts far enough behind the particle that the particle is initially in field-free vacuum, and that the particle's final energy is evaluated when the particle is once again in field-free vacuum (long after interaction with the pulse).
In the rest frame of the on-axis particle (of charge $q$ and mass $m$) traveling at its initial
($t=0$) z-directed velocity $v_0$ in the lab frame, the Lorentz force accelerates the particle
as

$$\frac{d(\gamma'\beta')}{dt'} = \frac{q}{mc} E_z(0,0,z,t) = A(z,t)\sin(\omega t - B(z)), \tag{1}$$

where the last equality makes an assumption about the on-axis form of the electric field
$E_z(x,y,z,t)$, the longitudinal component of the electric field in the laboratory frame.
$v = \beta c$ is the particle velocity and $\gamma \equiv 1/\sqrt{1-\beta^2}$, where $c$ is the speed of light in
vacuum. Primes indicate variables in the initial particle's rest frame, so the Lorentz
transform gives $z = \gamma_0(z'+v_0 t')$ and $t = \gamma_0(t'+(v_0/c^2)z')$, $\gamma_0$ being the initial $\gamma$. Let a
particle be considered relativistic if $|\gamma\beta| > g_b$, where $g_b$ is some reasonable value on the
order of 1. The mathematical statement of our hypothesis is

$$\overline{M}(\gamma_f/\gamma_0) >> 1 \rightarrow \overline{M}(M(|\gamma'(t')\beta'(t')|,t')) >> g_b, \tag{2}$$

where $M(f(x),x)$ is the maximum of $f$ over $x$, $\overline{M}(\cdot) \equiv M(M(\cdot,\psi_0),z(0))$. $\psi_0 \in [0,2\pi)$
is the carrier-envelope phase and $z(0) \in (-\infty,\infty)$ the initial particle position. Among the
most commonly-studied fields are the radially-polarized laser beam [11-16], for which
$E_z(0,0,z,t) = L_r \sin(\psi_r)\text{sech}(\xi_r/\xi_0)$, and the configuration of crossed Gaussian beams
[11,17-19], for which $E_z(0,0,z,t) = L_c X_c \sin(\psi_c)\text{sech}(\xi_c/\xi_0)$. $L_r \equiv \sqrt{8\eta_0 P/\pi}/(z_0(1+\zeta^2))$,
$\psi_r \equiv \omega t - kz + 2\tan^{-1}\zeta + \psi_0$, $\xi_r \equiv \omega t - k(z-z_i)$, $L_c \equiv \sin\theta\sqrt{8\eta_0 P/\pi}/(w_0(1+Z^2))$,

$X_c \equiv \exp(-b^2)$, $b^2 \equiv Z^2 \tan^2\theta/(\varepsilon^2(1+Z^2))$, $\psi_c \equiv \omega t - kz\cos\theta - Zb^2 + 2\tan^{-1}Z + \psi_0$,

$\xi_c \equiv \omega t - k(z\cos\theta - z_i)$; $\theta$ is the angle each beam makes with the axis in the crossed-beams scheme; $\lambda$ is the carrier wavelength; $z_0 \equiv \pi w_0^2/\lambda$; $k \equiv 2\pi/\lambda = \omega/c$; $w_0$ is proportional to the beam radius; $\zeta \equiv z/z_0$; $Z \equiv \zeta\cos\theta$; $\varepsilon \equiv w_0/z_0$; $\eta_0$ is the vacuum wave impedance; $z_i$ is the pulse's initial position (effectively $-\infty$); $\xi_0$ controls the pulse duration; $P$ is the total peak pulse power ($P/2$ peak power for each pulse in the crossed-beam scheme). Note that in either case, the field is of the form assumed in Eq. (1). Since we seek the boundary of negligible acceleration, where the particle energy does not change substantially according to our hypothesis, the particle approximately remains at its initial speed throughout ($v \approx v_0 \neq c\ \forall t$), so $z(t') \approx \gamma_0(\gamma_0 z(0) + v_0 t')$, $t(t') \approx \gamma_0(t' + v_0\gamma_0 z(0)/c^2)$. Eq. (1) may be solved as

$$\begin{aligned}
\gamma'(t')\beta'(t') &\approx \int_{t'(0)}^{t'} ds' A(z(s'),t(s'))\sin(\omega t(s') - B(z(s'))) \\
&\approx \int_{t'(0)}^{t'} ds' A(z(s'),t(s'))\sin\left(\omega\gamma_0 s' - \dot{B}(z(t'))\gamma_0 v_0 s' + const.\right) \\
&\approx \frac{-A(z(t'),t(t'))}{\omega\gamma_0(1-\dot{B}(z(t'))v_0/\omega)}\cos\left(\omega\gamma_0 t' - \dot{B}(z(t'))\gamma_0 v_0 t' + const.\right)
\end{aligned} \quad (3)$$

where $\dot{B} \equiv dB/dz$ and $t'(0) = -v_0\gamma_0 z(0)/c^2$. To arrive at the second line of Eq. (3), we Taylor-expanded B(z) and discarded higher order terms (assuming this is valid). We then integrated by parts and noted that $A(z,t) \equiv a(z)\text{sech}(\xi/\xi_0)$, the product of the beam and pulse envelopes, varies slowly compared with the carrier sinusoid to arrive at the third line. We then insert Eq. (3) into Eq. (2) to get

$$\overline{M}\left(\frac{\gamma_f}{\gamma_0}\right) \gg 1 \to M\left(\left|\frac{a(z)}{\omega\gamma_0\left(1-\dot{B}(z)v_0/\omega\right)}\right|, z\right) \gg g_b, \tag{4}$$

where the optimizations over $\psi_0$ and $z(0)$ allowed us to set the sinusoid and the pulse envelope respectively in Eq. (3) to their maximum value of 1. It is straightforward to verify, by taking first and second derivatives, that $M$'s first argument in Eq. (4) is maximized at $z=0$ (ignoring singularities) for both the radially-polarized laser beam and the crossed-beams cases. For the radially-polarized laser beam, Eq. (4) becomes

$$\overline{M}\left(\frac{\gamma_f}{\gamma_0}\right) \gg 1 \to P \gg \gamma_0^2\left(1-\beta_0\left(1-\varepsilon^2\right)\right)^2 \left(\frac{g_b mc^2}{q\varepsilon^2}\right)^2 \frac{\pi}{2\eta_0}. \tag{5}$$

Setting $\beta_0 = 0$ in Eq. (5) gives the threshold power for an initially stationary electron obtained by a different procedure in [13]. For the crossed-beams configuration, we have

$$\overline{M}\left(\frac{\gamma_f}{\gamma_0}\right) \gg 1 \to P \gg \gamma_0^2\left(1-\beta_0\cos\theta\left(1-\varepsilon^2\right)\right)^2 \left(\frac{g_b mc^2}{q\varepsilon\sin\theta}\right)^2 \frac{\pi}{2\eta_0}. \tag{6}$$

Eqs. (5) and (6) (and (4), of which they are special cases) are useful analytical approximations of Eq. (2), but only when the assumptions we have made in obtaining them are valid. For instance, when $\gamma_0 \gg 1$ in the case of Eq. (5), or $\gamma_0 \gg 1$ and $\cos\theta \approx 1$ in the case of Eq. (6), the width of $a(z(t'))$ in $t'$ may be comparable to the period of the sinusoidal carrier in $t'$, contrary to our assumption in Eq. (3) that $A$ varies slowly with respect to the sinusoidal carrier in the particle's frame. In such cases, one would expect Eq. (4) to be a relatively poor estimate of Eq. (2). To compare our theory with the results of exact numerical simulations, we solve Eq. (1) and $dz/dt = v$ using a fourth-order Runge-Kutta algorithm, optimizing for energy gain over $\psi_0$ - $z(0)$ space in various two-dimensional parameter sweeps. These are plotted in Figs. 1 and 2, where the corresponding acceleration threshold (setting $g_b = 1$) hypothesized in Eq. (2) as well as the analytical approximations obtained from Eq. (4) are also displayed. In all plots, Eq. (2) approximates the threshold of substantial acceleration with high accuracy, and Eq. (4) is a fair approximation of Eq. (2) most of the time.

Fig. 1 shows several parameter sweeps for electron acceleration by a pulsed radially-polarized beam. For many plots, we have chosen $\xi_0 = 13.37$ because it corresponds to a FWHM pulse duration of 10 fs for $\lambda = 0.8 \mu m$. Fig. 2 shows parameter sweeps for the crossed-beams configuration. We note from Figs. 1(a)-(b) and 2(a)-(b) that the threshold power does not decrease indefinitely with increasing $K(0)$, but scales as $\gamma_0^2$ when $\beta_0 \to 1$ (one intuitively expects this behavior from the fact that the theoretical energy gain limit scales as $P^{1/2}$ [13,16]). This implies that, unlike conventional accelerators, effectively-unbounded linear particle accelerators cannot be cascaded indefinitely for greater gain: the energy of an output particle is ultimately limited by the peak power of the strongest laser in the cascade.

In conclusion, our simulations support our hypothesis that substantial net linear acceleration is contingent on the accelerating field's ability to bring the particle to a relativistic energy in its initial rest frame during the interaction, at least for the types of beams and range of parameters considered in this paper. In the process, we have derived a general formula for the acceleration threshold, which is practically useful as a guide to the laser intensities that unbounded linear acceleration requires. The fact that a relativistic particle can be further accelerated by unbounded linear acceleration is important because this enables the injection of a relativistic particle beam, which is more resistant to space-charge effects than a non-relativistic beam is. Although we have illustrated our theory with electron acceleration by a radially-polarized laser beam and the crossed-beams configuration, our theory may be readily extended to any other unbounded linear acceleration scheme that can be described by an equation of the general form in Eq. (1). Future work will concern the optimization of multi-particle vacuum-based linear acceleration schemes.

Acknowledgments

This work was supported by the National Science Foundation (NSF) grant NSF-018899-001 and the Agency for Science, Technology and Research (A*STAR), Singapore.

# Figures

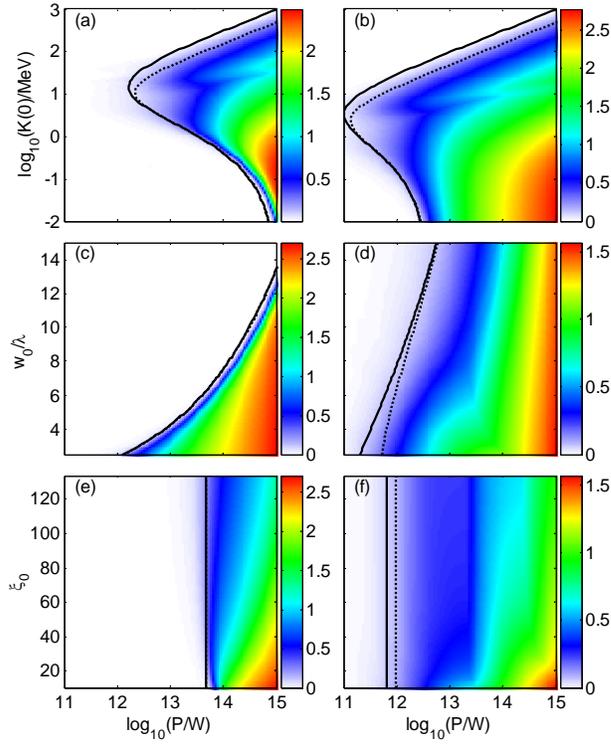

Fig. 1: Color maps of $\log_{10}(\gamma_f/\gamma_0)$ ($\gamma_f/\gamma_0$ being ratio of final to initial particle energy) for linear electron acceleration by a pulsed radially-polarized laser beam as a function of various parameters: initial kinetic energy $K(0)$ and peak pulse power $P$ for normalized pulse duration $\xi_0 = 13.37$ and (a) normalized beam radius $w_0/\lambda = 10$, (b) $w_0/\lambda = 2.5$; $w_0/\lambda$ and $P$ for $\xi_0 = 13.37$ and (c) $K(0) = 0.1\,\text{MeV}$, (d) $K(0) = 10\,\text{MeV}$; $\xi_0$ and $P$ for $w_0/\lambda = 6.25$ and (e) $K(0) = 0.1\,\text{MeV}$, (f) $K(0) = 10\,\text{MeV}$. Solid black lines demarcate the acceleration threshold predicted by our hypothesis. Dotted black lines correspond to the analytical approximation of this boundary.

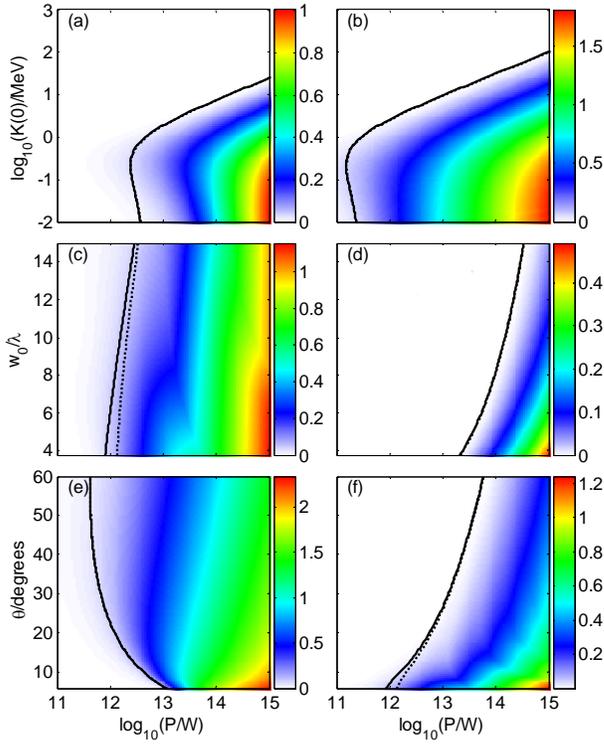

Fig. 2: Color maps of $\log_{10}(\gamma_f/\gamma_0)$ ($\gamma_f/\gamma_0$ being ratio of final to initial particle energy) for linear electron acceleration by the crossed-beams configuration as a function of various parameters: initial kinetic energy $K(0)$ and total peak pulse power $P$ for normalized pulse duration $\xi_0 = 13.37$, (a) normalized beam radii $w_0/\lambda = 15$ and crossing angle $\theta = 45^0$, (b) $w_0/\lambda = 3.75$ and $\theta = 45^0$; $w_0/\lambda$ and $P$ for $\xi_0 = 13.37$, (c) $K(0) = 10\,\text{MeV}$ and $\theta = 2.5\lambda/\pi w_0$, (d) $K(0) = 10\,\text{MeV}$ and $\theta = 60^0$; $\theta$ and $P$ for $\xi_0 = 13.37$, (e) $w_0/\lambda = 6.25$ and $K(0) = 0.1\,\text{MeV}$, (f) $w_0/\lambda = 6.25$ and $K(0) = 10\,\text{MeV}$. Solid black lines demarcate the acceleration threshold predicted by our hypothesis. Dotted black lines correspond to the analytical approximation of this boundary.

[1]Department of Electrical Engineering and Computer Science and Research Laboratory of Electronics, Massachusetts Institute of Technology, 77 Massachusetts Avenue, Cambridge, MA, 02139, USA.

[2]Center for Free-Electron Laser Science, DESY, and Department of Physics, University of Hamburg, Notkestraße 85, D-22607 Hamburg, Germany.

[a)] Electronic address: ljwong@mit.edu